\documentclass{aa}

\usepackage{graphicx}
\usepackage{psfrag}

\def\Div{\mathop{\hbox{div}}\nolimits}

\newcommand{\beq}{\begin{equation}}
\newcommand{\eeq}{\end{equation}}
\newcommand{\eeqn}[1]{\label{#1}\end{equation}}
\newcommand{\greq}{\begin{equation}\left\{ \begin{array}{l}}
\newcommand{\egreq}{\end{array}\right. \end{equation}}
\newcommand{\egreqn}[1]{\end{array}\right. \label{#1}\end{equation}}
\newcommand{\eq}[1]{(\ref{#1})}
\newcommand{\ltex}[1]{\quad \hbox{#1} \quad}
\newcommand{\lp}{ \left(}
\newcommand{\rp}{ \right)}
\newcommand{\na}{ \vec{\nabla} }
\newcommand{\disp}[1]{\displaystyle #1}
\newcommand{\dz}[1]{\frac{\partial  #1}{\partial z}}
\newcommand{\del}{\nabla}
\newcommand{\delad}{\nabla_{\hbox{\scriptsize ad}}}

\newcommand{\mad}{m_{\hbox{\scriptsize ad}}}
\newcommand{\dnz}[1]{\frac{d  #1}{dz}}
\newcommand{\vxi}{\vec{\xi}}
\newcommand{\fc}{F_{\hbox{\scriptsize conv}}}
\newcommand{\fb}{F_{\hbox{\scriptsize bot}}}
\newcommand{\fr}{F_{\hbox{\scriptsize rad}}}
\newcommand{\BV}{Brunt-V\"ais\"al\"a\ }

\begin{document}

\title{Spectrum and amplitudes of internal gravity \\
waves excited by penetrative convection \\
in solar-type stars}
\titlerunning{IGW excitation by penetrative convection}

\author{Boris Dintrans\inst{1} \and Axel Brandenburg\inst{2} \and
$\hbox{\AA}$ke Nordlund\inst{3} \and Robert F. Stein\inst{4}}

\offprints{boris.dintrans@obs-mip.fr}

\institute{Observatoire
Midi-Pyr\'en\'ees, UMR5572, Universit\'e Paul Sabatier et CNRS,
F-31400 Toulouse, France \and
NORDITA, Blegdamsvej 17, DK-2100 Copenhagen \O, Denmark \and
Niels Bohr Institute, Juliane Maries Vej 30, DK-2100 Copenhagen \O,
Denmark \and Department of Physics and Astronomy, Michigan State
University, East Lansing, MI 48824, U.S.A.}

\date{\today}

\abstract{
The excitation of internal gravity waves by penetrative convective
plumes is investigated using 2-D direct simulations of compressible
convection. The wave generation is quantitatively studied from the
linear response of the radiative zone to the plumes penetration,
using projections onto the g-modes solutions of the associated linear
eigenvalue problem for the perturbations. This allows an accurate
determination of both the spectrum and amplitudes of the stochastically
excited modes. Using time-frequency diagrams of the mode amplitudes,
we then show that the lifetime of a mode is around twice its period
and that during times of significant excitation up to 40\% of
the total kinetic energy may be contained into g-modes.
 }

\maketitle

\section{Introduction}
\label{intro}

Although their detection in the spectrum of solar oscillations has been
not clearly confirmed (e.g., Turck-Chi\`eze et al.\ 2004; Gabriel et
al.\ 2002),
internal gravity waves (hereafter IGWs) propagating in the radiative
zones of late-type stars have recently been invoked in
attempts to explain the Li abundance of cool stars and the rigid
rotation of their radiative interiors.

The former problem is also referred to as the Li-dip problem and concerns
the dependence of the lithium abundance on the spectral type for some
main-sequence stars. Models based on the
extension of the surface convection zone down to the nuclear burning region
(Iben 1965) and models which take into account the transport of Li both by
meridional circulation and shear-induced turbulence (Talon \& Charbonnel
1998) are indeed not quite satisfactory to reproduce the Li-dip.
The latter problem concerning
the rigidity of the Sun's radiative interior is most clearly revealed by
helioseismology (Brown et al.\ 1989). Both rotation-induced
turbulent diffusion and wind-driven meridional circulation fail to
redistribute enough angular momentum over the lifetime of the Sun to
rotate rigidly (Zahn et al.\ 1997).
Likewise, the hypothesis of a large-scale poloidal magnetic
field leads to problems, because it may transmit under certain
circumstances the differential rotation
of the convection zone to the core (owing to Ferraro's law;
e.g., MacGregor \& Charbonneau 1999).

Zahn (1994) showed that the two problems are coupled and that they should to be
explained by a single model. Being inspired by meteorological studies of
the wave transport taking place in the Earth stratosphere (Bretherton
1969; Alexander \& Pfister 1995), Schatzman (1993) was the first to
propose internal gravity waves as an efficient transport mechanism in stellar
radiative zones. Later, this idea was tested by many authors (Zahn et al.\
1997; Kumar \& Quataert 1997; Kumar et al.\ 1999; Talon, Kumar \& Zahn
2002) who showed that
internal gravity waves transport momentum on a rather short timescale such that the
rotation of the solar core becomes nearly uniform. A remaining problem is the
excitation of these deep gravity waves since, unlike pulsating white
dwarfs, a $\kappa$-mechanism based on hydrogen and helium ionization
zones is not applicable here.

The most wide-spread excitation model is based on penetrative
convection from neighboring convection zones. Strong downward plumes
are known to extend a substantial distance into the adjacent stable
zones so that internal gravity waves can be randomly generated. 2-D
and 3-D direct numerical simulations of superposed stable and unstable
layers have confirmed this scenario since gravity-like waves have been
well observed in stable regions (Hurlburt et al.\ 1986, 1994; Brandenburg
et al.\ 1996; Brummell et al.\ 2002).

The aim of this paper is to investigate in detail the excitation of
IGWs by overshooting in an high-resolution 2-D
hydrodynamical simulation of a three-layer model, consisting of a convective
zone (hereafter CZ) embedded between an upper cooling zone and a lower
stably stratified radiative zone (hereafter RZ). In all
previous
numerical studies,
gravity waves have been detected using Fourier's analysis such as, for
example, the
$(k,\ \omega)$-diagram. However, the link with the eigenmodes of stable
regions was not investigated. In particular, it has not clearly been
demonstrated that the excited waves really correspond to gravity waves
since the only criterion was the Brunt-V\"ais\"al\"a\ frequency
as an upper bound.

In this work, g-modes are rigorously measured using the method that
we have presented and tested on the gravity mode oscillations of an
isothermal atmosphere in Dintrans \& Brandenburg (2004, hereafter Paper
I). Our method mainly relies on the projection of the velocity field of
the simulation onto the basis shaped from the solutions of the associated
linear eigenvalue problem for the perturbations; i.e., the theoretical
g-modes of the RZ. Hence the mode amplitudes are simply given by the
time-dependent basis coefficients, which allows a quantitative study of
the excitation mechanism. In other words, we investigate the generation of
IGWs by penetrative convection from the linear response of the RZ to
this penetration.

We begin by presenting our hydrodynamical 2-D model consisting in three
superposed polytropic layers, and give some details on the code we use to
solve it numerically (Sect.\ \ref{themodel}). We then give the main
properties of the obtained simulation of penetrative convection and show
that the classical
detection method of gravity mode oscillations based on Fourier's
transforms in both space and time fails to give reliable results on this
problem (Sect.\ \ref{nature}). We then introduce our detection technique
based on the anelastic subspace and apply it to find the properties of
the g-modes propagating in the numerical simulation (Sect.\
\ref{results}).  Finally, we conclude in Sect.\ \ref{conclu} by giving
some outlooks of this work.

\section{The hydrodynamical model}
\label{themodel}

\subsection{The basic equations}
\label{basic}

\begin{figure}
\psfrag{xaxis}{$x$}
\psfrag{zaxis}{$z$}
\psfrag{x1}{$x_1$}
\psfrag{x4}{$x_4$}
\psfrag{z1}{$z_1$}
\psfrag{z2}{$z_2$}
\psfrag{z3}{$z_3$}
\psfrag{z4}{$z_4$}
\psfrag{upper}{stable cooling zone with index $m_1$}
\psfrag{middle}{unstable convection zone with index $m_2$}
\psfrag{lower}{stable radiative zone with index $m_3$}
\centerline{\includegraphics[width=7cm]{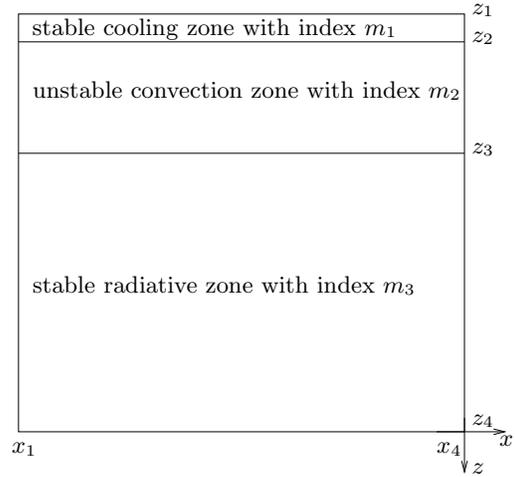}}

  \caption{Geometry of the computational domain.}

\label{geometry}
\end{figure}

We adopt Cartesian coordinates $(x,z)$ where $x$ denotes the horizontal
direction and $z$ is depth pointing downward as the gravity
$\vec{g}$. Our system is
composed of a convection zone of depth $d=z_3-z_2$, embedded
between two stable layers (Figure \ref{geometry}). We assume that the
gas is monatomic and perfect, so its equation of state is given by

\[
p = (\gamma -1) \rho e \ltex{with} \gamma=c_p/c_v=5/3,
\]
where $p$ is the pressure, $\rho$ the density, $e$ the internal energy,
and $\gamma$ is the ratio of specific heats $c_p$ and $c_v$.

We solve the following set of hydrodynamical equations (conservation of
mass, momentum and energy):

\greq
\displaystyle \frac{D \ln \rho}{Dt} = -\Div \vec{u}, \\ \\
\displaystyle \frac{D \vec{u}}{Dt} = - (\gamma-1) \lp \na e + e \na
\ln \rho \rp + \vec{g} + \frac{2}{\rho} \na \cdot (\rho \nu
\vec{\sf S}), \\ \\
\disp \frac{De}{Dt} = - (\gamma-1) e \Div \vec{u} +
\frac{1}{\rho} \na \cdot ({\cal K} \na e) + 2\nu
\vec{\sf S}^2 + {\cal Q},
\end{array} \right.
\eeqn{full}
where $\vec{u}$ is the velocity and
$D/Dt = \partial / \partial t + \vec{u} \cdot \na$ is
the total derivative. In addition, $\nu$
denotes the {\it constant} kinematic viscosity\footnote{In Hurlburt et
al.\ (1986, 1994), Cattaneo et al.\ (1991) or Bogdan et al.\ (1993), it is
the dynamical viscosity $\mu = \rho \nu$ which is constant.
However, the kinematic viscosity becomes
very large as the density tends to zero leading to a highly viscous surface
layer.} and ${\cal K}=K/c_v$ the radiative conductivity divided by
$c_v$\footnote{Hereafter, we will simply refer to ${\cal K}$ as the
radiative conductivity.}. To reproduce the radiative cooling taking
place at a star's surface, we add a cooling term $\cal Q$ in the energy
equation given by

\beq
{\cal Q} = - \left(e - e_{\hbox{\scriptsize top}}\right){\tau}^{-1}(z),
\eeqn{cooling}
where $\tau^{-1}(z)$ is the cooling rate profile which is set equal to zero
everywhere except close to the surface, i.e.\ $\tau^{-1}(z) \ne 0$ only for
$z=[z_1,z_2]$. Because of this efficient cooling, the surface layer tends
to be isothermal and the internal energy $e$ is almost constant
there. Finally, $\vec{\sf S}$ denotes the traceless strain
tensor given by

\[
{\sf S}_{ij} = \frac{1}{2}\left(\frac{\partial u_i}{\partial x_j} +
\frac{\partial u_j}{\partial x_i}\right) - \frac{1}{3} \delta_{ij} \na
\cdot \vec{u}.
\]

\subsection{Boundary conditions and the initial setup}

We assume that $\ln \rho, \vec{u}$ and $e$ are periodic in
the horizontal direction and adopt the following conditions at the upper
and lower boundaries:

\greq
\disp \dz{u} = \dz{w} = w = 0 \ltex{at} z=z_1,z_4; \\ \\
\disp e=e_{\hbox{\scriptsize top}} \ltex{at} z=z_1 \ltex{and}
\dz{e} \ltex{fixed at} z=z_4,
\egreq
where $u$ and $w$ correspond to the horizontal and vertical velocities,
respectively.

Following Brandenburg et al.\ (1996),
we choose the depth of the unstable layer $d$ as the
unit of length, $(d/g)^{1/2}$ as the unit of time and the initial value
$\rho_{\hbox{\scriptsize bot}}$ of the density at the
bottom of the convection zone (hereafter BCZ) as the
unit of density [velocities are thus measured in units of $\sqrt{gd}$,
i.e.\ the free-fall velocity of the unstable layer divided by $\sqrt{2}$,
and fluxes in units of $\rho_{\hbox{\scriptsize bot}} (gd)^{3/2}$].
Finally, the dimensionless gravity is set equal to unity in the whole
domain, i.e.\ $g=1$ everywhere.

The initial state is computed using polytropes in hydrostatic and
radiative equilibrium (e.g., Hurlburt et al.\ 1986).
In Appendix \ref{init} we give
the details of the initial setup, in particular
the mixing length solution in the calculation of
the CZ stratification, which helps to accelerate the numerical convergence
towards the thermally relaxed state.

\begin{figure}
\centerline{\includegraphics[width=9cm]{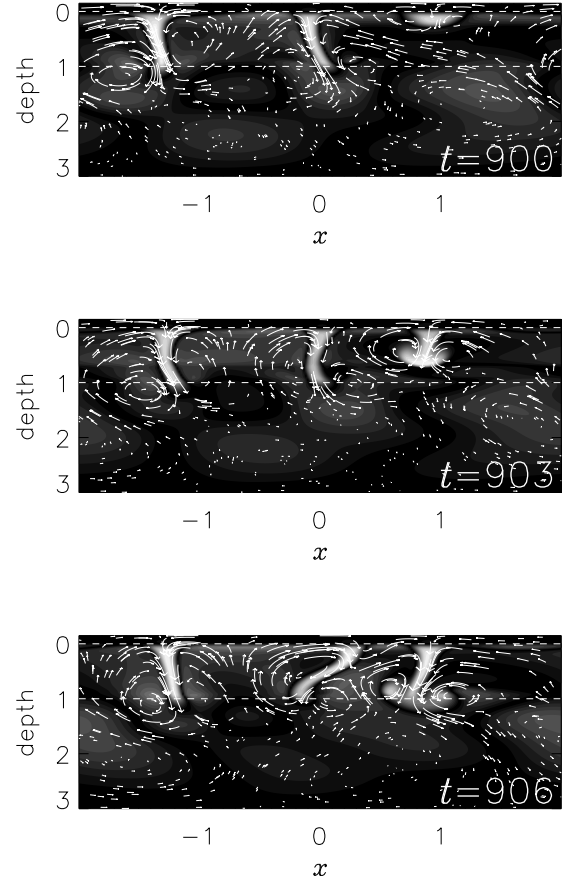}}

\caption{Snapshots of the velocity field superimposed on a grey scale
representation of
the internal energy fluctuations (i.e.\ or temperature) at three different
times $t=[900,\ 903,\ 906]$. The horizontal dashed
lines delimit the convection zone $z=[0,1]$ located above the radiative
zone $z=[1,3]$.}

\label{example}
\end{figure}

\subsection{The numerical method}

We use the hydrodynamical code described in Nordlund \& Stein (1990) to
advance the fully nonlinear set of Eqs. \eq{full}. In this code, spatial
derivatives are computed using sixth order compact finite differences
(Lele 1992) whereas the time advance is performed using a third order
explicit Hyman scheme (Hyman 1979). Corresponding timesteps are usually
$25\%$ of the Courant-Friedrich-Levy timestep defined by

\[
\delta t_{\hbox{\scriptsize CFL}} = \Delta z / \max (c_s, |\vec{u}|,
\alpha \nu/\Delta z,\alpha \chi /\Delta z),
\]
where $c_s$ is the sound speed, $\chi={\cal K}/(\gamma \rho)$ the
radiative diffusivity and $\alpha \simeq 4$ an empirical factor
determining the length of the diffusive timestep. In practice, it is the
radiative diffusion term which limits the timestep due to both the high
vertical resolution and large radiative diffusivities of radiative
zones.

Contrary to acoustic modes, gravity modes correspond to waves with
long periods, of about twenty time units in our numerical simulations,
typically. Hence, once achieved the thermal relaxation of the radiative
zone, we still need to integrate the dynamical equations over very long
times to capture so much periods as possible (one typically needs runs
as long as one thousand time units). We then chose to concentrate on a
particular 2-D simulation with an high spatial resolution $256\times 512$
(i.e.\ 256 mesh points in the horizontal and 512 points in the vertical)
and an aspect ratio $A\equiv L_x/d=4$ ($L_x$ being the horizontal extent
of the computational domain). The main properties of this simulation
are the following:

\greq
m_1 = -0.9,\quad m_2=-0.8,\quad m_3 = 3,\quad
e_{\hbox{\scriptsize top}}=0.3, \\ \\
F_{\hbox{\scriptsize tot}}=5\times10^{-3},\quad \nu = 5\times 10^{-3},
\\ \\
\hbox{Pr}_2=12.5,\quad \hbox{Pr}_3 = 0.625,
\egreq
where $\hbox{Pr}=\nu/\chi$ denotes the Prandtl number. The corresponding
Rayleigh number of the unstable layer is $\hbox{Ra} \simeq 8.5\times
10^5$, using the definition in Gough et al.\ (1976). The kinematic
viscosity $\nu$ cannot be too small for a given resolution as it should
satisfy $d / \Delta z \sim \hbox{Re}^{3/4}$, where $\Delta z$ is the smallest
mesh interval in the vertical direction and Re is the Reynolds number
(e.g., Landau \& Lifshitz 1980). The value we chose is a reasonably
``safe'' one as this
is around fifty times larger than the minimum viscosity $\nu_{\min}
\sim 10^{-4}$ imposed by both the resolution $256\times512$ and the mean
Reynolds number of our simulation.

\begin{figure}
\centerline{\includegraphics[width=9cm]{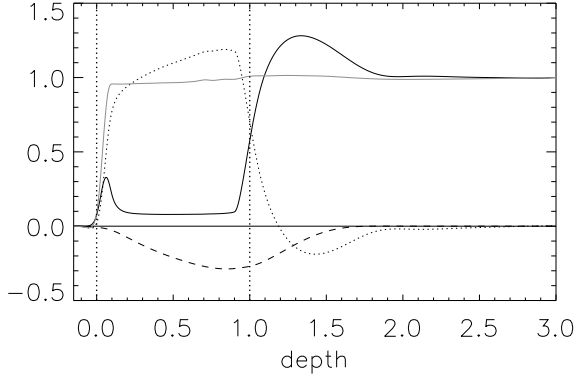}}

\caption{Normalized vertical profiles of the radiative (solid
dark line), convective (dashed line) and kinetic (dot-dashed line)
fluxes, with their sum (solid grey line). The
vertical dotted lines denote the CZ limits.}

\label{fluxes}
\end{figure}

\begin{figure}
\centerline{\includegraphics[width=9cm]{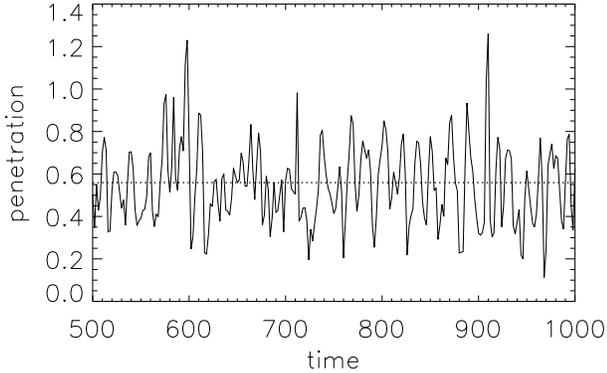}}

\caption{Evolution with time of the penetration extent $\Delta$,
with its time-averaged value
$\left<\Delta\right>\simeq 0.56$
denoted by the horizontal dotted line. }

\label{penetration}
\end{figure}

\begin{figure}
\centerline{\includegraphics[width=8cm]{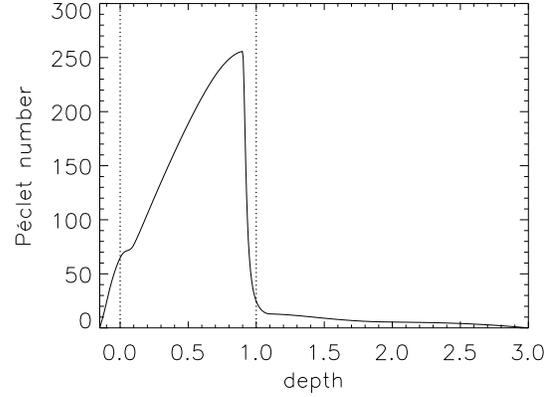}}

\caption{Mean vertical profile of the P\'eclet number,
where dotted lines mark the CZ limits.}

\label{peclet}
\end{figure}

\section{Nature of the penetrative convection and the IGW detection
problem}
\label{nature}

Since the pioneering numerical simulations of Hurlburt et al.\ (1986), the
general features of penetrative compressible convection are well
known and we are simply giving here the main properties of such a flow.

Figures \ref{example} and \ref{fluxes} represent typical asymmetrical
patterns and mean vertical radiative, convective and kinetic fluxes that
we obtain in our numerical simulation of
compressible convection with penetration, the fluxes being computed from
their usual definition (e.g., Hurlburt et al.\ 1986)

\greq
\disp F_{\hbox{\scriptsize rad}} = K \dnz{} \langle
\overline{T}\rangle_t,\\ \\
\disp F_{\hbox{\scriptsize conv}} = -c_p \langle \overline{\rho w
T'}\rangle_t,\\ \\
\disp F_{\hbox{\scriptsize kin}} = - \frac{1}{2} \langle \overline{w \rho
(u^2 + w^2)}\rangle_t,
\egreq
where the overbars and brackets denote the horizontal and temporal
means, respectively, and $T'$ is the temperature fluctuation
about the horizontal mean.
Compressibility effects
destroy the usual symmetric velocity field observed in Boussinesq
convection as downdrafts become stronger and more concentrated than the broader
upward flow (Graham 1975; Hurlburt et al.\ 1984). The main consequence is that
some localized downward-directed narrow plumes appear which transport a
significant kinetic flux in this direction (around 30\% of the total
flux at the BCZ in this simulation; see the dot-dashed line in
Fig.\ \ref{fluxes}).

The presence of a lower stable layer below the convection zone allows
these
downward plumes to penetrate some distance into the RZ. This convective
penetration has been theoretically investigated in the astrophysical
context by Zahn (1991) who showed that it strongly depends on the
value of the local P\'eclet number $\hbox{Pe} = w L/\chi$ ($w$ and $L$
being the typical vertical velocity and size of those motions,
respectively).

Following Brummell et al.\ (2002), we define the penetration extent
$\Delta$ as the vertical distance from the BCZ where the
horizontally-averaged kinetic flux decreases to 1\% of its maximum value
and Fig.\ \ref{penetration} shows an example of its evolution with
time. The averaged penetration is $\left<\Delta\right>\simeq 0.56$, that is,
of order the half-size of the CZ, which is typical of hydrodynamical
simulations at low P\'eclet numbers. Indeed, $\hbox{Pe}$ falls down
very rapidly just below the CZ due to the large radiative diffusivity of
the radiative zone.  In polytropic models, the radiative diffusivity is
proportional to $1+m$, where $m$ is the polytropic index,
see Eq.~(\ref{rad0}), so a convective blob experiences
a jump in the P\'eclet number when it crosses the interface between the
CZ and RZ zones

\beq
\frac{\hbox{Pe}_3}{\hbox{Pe}_2} \sim \frac{\chi_2}{\chi_3} \propto
\frac{1+m_2}{1+m_3}.
\eeqn{pe_jump}
Figure \ref{peclet} shows the vertical profile of the P\'eclet
number. The estimate of the P\'eclet jump across
the CZ-RZ interface using Eq.\eq{pe_jump} gives $\hbox{Pe}_3/\hbox{Pe}_2
\simeq 0.05$ or $\hbox{Pe}_3 \simeq \hbox{Pe}_2 / 20$, which is
effectively the one in Fig.\ \ref{pe_jump}, where $\hbox{Pe}_3 / \hbox{Pe}_2
\simeq 13/250=0.05$. As a result, the cold blob thermalizes very rapidly
in the radiative zone and loses its identity compared to its environment:
the buoyancy braking thus disappears and allows the blob to continue by
inertia in the stable zone, leading to the observed large penetration.
(In solar-type stars the relevant extent is actually much smaller,
because here the thermal conductivity is still too large and therefore
the overall fluxes, and hence the amount of flux is too large.)

\begin{figure}
\centerline{\includegraphics[width=9cm]{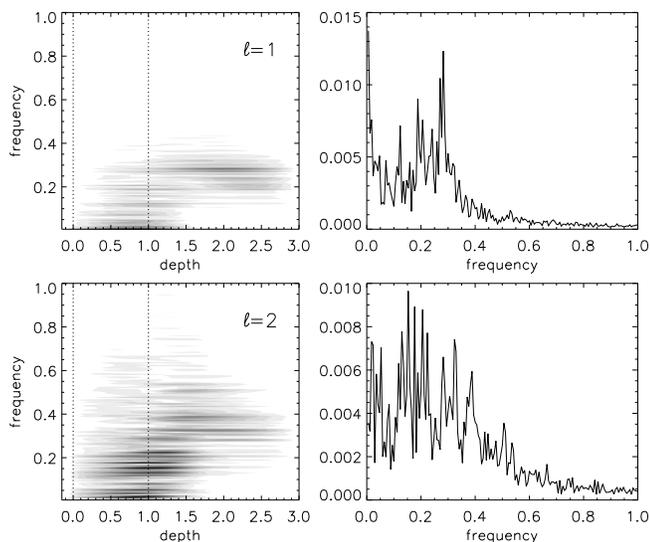}}

\caption{{\it Left:} Gray scale representations in the $(z,\omega)$-plane
of the low-frequency part of the temporal power spectra for two degrees
$\ell=[1,2]$. {\it Right:} the resulting spectra after
integrating over depth.}

\label{classic}
\end{figure}

\subsection{Finding the g-modes: the problem of the random excitation}
\label{metclass}

As discussed in Paper I,
the classical technique to detect the g-modes
propagating in an hydrodynamical simulation consists first of taking
horizontal Fourier transforms of the vertical mass flux $\rho w (x,z,t)$
for every time step, to get $\widehat{\rho w}(\ell,z,t)$\footnote{Here
and in the following, we define the horizontal wavenumber as $k_x =
(2\pi/L_x)\ \ell=(\pi/2)\ \ell$, where $\ell$ denotes the mode degree
and is a non-zero positive integer, $\ell = [1,\ 2, \dots]$, as purely 
radial g-modes cannot exist (e.g., Turner 1973).}.
Second, one computes power spectra
for the
individual time series $\widehat{\rho w}(\ell,z,t)$, to get $\widehat{\rho
w}(\ell,z,\omega)$, and plots the resulting power spectra at a given
degree $\ell$ in a $(z,\omega)$-plane to highlight the ``shark fin''
peaks corresponding to the eigenmodes (e.g., Fig.\ 5 in Paper I).

Figure \ref{classic} shows the result of applying this method to the
detection of $\ell = 1$, $\ell = 2$ and $\ell = 3$ g-modes propagating in
the simulation. Low-frequency peaks appear in the RZ ($1\le z \le
3$), mainly in the $\ell=1$ diagram, but these peaks are not as well
defined as in Paper~I. Indeed, we have focused in the previous paper
on the simpler case of the g-modes of an isothermal atmosphere where IGWs
were excited by the vertical free oscillations of a cold bubble: once
emitted, each global mode ``stayed in the box'' during a time that depends
on the efficiency of the dissipation, which was mainly due to a weak
viscosity. As a consequence, large-scale nonradial g-modes survived
over long times and were very well visible in the temporal spectra,
both in the $(z,\omega)$-plane and in its depth-integrated representation
(see Fig.\ 3 in Paper~I).

The situation is clearly different with penetrative convection as we have now
to deal with a random excitation of IGWs. Indeed, strong downward plumes
are not stationary structures: they are born in the upper layers of the
convection zone, are accelerated by the Archimedes force during their CZ
crossing and, finally, end their life in the overshoot region where
they transfer a large amount of their stored kinetic energy to the
stably stratified medium, resulting in an internal gravity wave field. As the
birth of these plumes is random,
the resulting forcing of the RZ wave field is itself a
random one, exactly as an hammer which randomly strikes the upper part
of the RZ.  We then understand why the detection of these IGWs is
more difficult than the simpler case of free bubble oscillations:
once emitted by a penetrating plume, a global mode first propagates in
the radiative zone while being subjected to the viscous and radiative
dissipations. However, if a second plume arrives shorter afterwards, the mode
pattern may be destroyed, resulting in partial interference and the
corresponding frequency peak may disappear from the spectrum. In other words,
the lifetime of a mode is not simply related to the diffusive processes
but also to the frequency of plume penetration (as we will show in
Sect.\ \ref{evol}).

With this in mind, it is also clear that the classical detection
method based on successive Fourier transforms both in space and
time of the vertical mass flux is not well adapted to detecting IGWs in
simulations with penetrative convection. The temporal Fourier transforms
are computed
over the whole simulation while IGWs are only present during a short
time interval. This results in a mixing of wave
events with non-wavy turbulent events such that the spectra lack
a well defined frequency (Fig.\ \ref{classic}, right).

\section{Results}
\label{results}

\subsection{IGW identification using the anelastic subspace}

\begin{figure}
\centerline{\includegraphics[width=9cm]{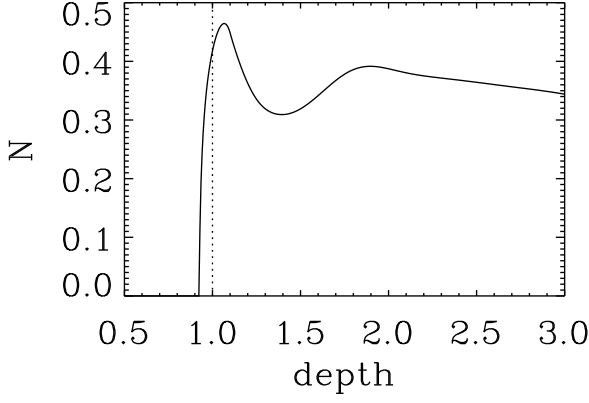}}

\caption{Mean vertical profile of the \BV frequency $N$ in
the RZ, the dotted line denoting the BCZ.}
\label{bv}

\end{figure}

We have presented in Paper I a new detection method of IGWs
in hydrodynamical simulations which is based on the anelastic
subspace and we will give here only its main
stages. The idea is to project
the simulated velocity field onto the basis built
with the anelastic eigenvectors which are solutions of the following
linear problem for the adiabatic perturbations (see also Dintrans \&
Rieutord 2001; Rieutord \& Dintrans 2002)

\greq
\disp \omega^2 \lp \xi_z - \frac{1}{k_x} \dnz{\xi_x} \rp = N^2
\xi_z, \\ \\
\disp -k_x \xi_x + \dnz{\xi_z} + \dnz{\ln \rho_0} \xi_z = 0, \\ \\
\xi_z = 0 \ltex{for} z=z_1,z_4,
\egreqn{anel}
where $\vxi=(\xi_x,\xi_z)$ denotes the Lagrangian displacement vector (the
velocity field being $\vec{u} = \hbox{i} \omega \vxi$),
$\rho_0$ the mean density (i.e., $\rho_0 =\langle
\overline{\rho}\rangle_t$) and $N^2$ the
square of the Brunt-V\"ais\"al\"a
frequency given by

\[
N^2 = g \left[ {1\over\gamma} \dnz{\ln e_0} - \lp 1-{1\over\gamma}\rp
\dnz{\ln \rho_0} \right].
\]
To derive the anelastic subset \eq{anel}, we first filtered out the
acoustic waves in the governing equations for the linear perturbations,
second assumed that the time dependence of normal modes is of the form
$\exp (\hbox{i}\omega t)$ and, finally, that their horizontal dependence
follows from the imposed periodicity in this direction as

\[
\xi_x \propto \cos (k_x x) \ltex{and} \xi_z \propto \sin (k_x x).
\]
The coupled differential
equations \eq{anel} form a generalized eigenvalue problem of the form

\beq
{\cal M}_A \vec{\psi}_{\ell n} = \omega^2_{\ell n} {\cal M}_B
\vec{\psi}_{\ell n},
\eeqn{eigenv}
where $\vec{\psi}_{\ell n} = (\xi_x,\xi_z)^T$ is the eigenvector of degree
$\ell$ and radial order $n$ associated with the eigenvalue $\omega^2_{\ell
n}$, while ${\cal M}_A$ and ${\cal M}_B$ denote two differential
operators.

\begin{figure}
\centerline{\includegraphics[width=9cm]{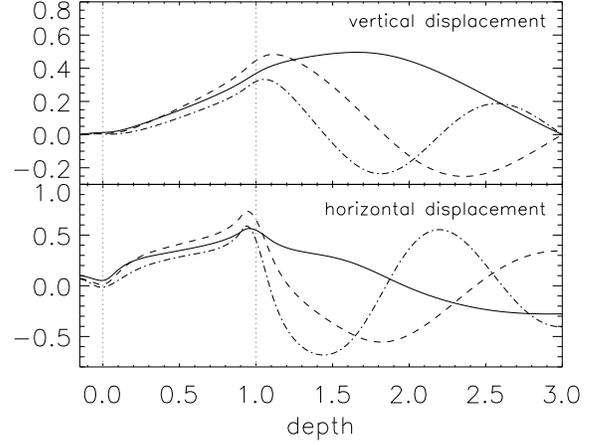}}

\caption{First three anelastic eigenvectors $\vec{\psi}_{\ell n}=
(\xi_x,\xi_z)^T$ at $\ell = 1$ and radial orders $n=0$ (solid line),
$n=1$ (dotted line) and $n=2$ (dashed-dotted line). Upper and
lower panels show the vertical and horizontal displacements,
respectively. Eigenvectors have been normalized by imposing
$\int_{z_1}^{z_4} \rho_0 |\vec{\psi}_{\ell n}|^2\ \rm{d}z = 1$.}

\label{vecp}
\end{figure}

The profile of the \BV frequency $N$ in the radiative zone is
shown in Fig.\ \ref{bv}. The g-modes spectrum is bounded by its
maximum value, $\max (N)\simeq 0.46$, while the typical frequency of
a low-degree and low-order g-mode is given by $\omega_{\ell n} \sim
\overline{N} \simeq 0.35$, hence a nondimensional period $T_{\ell
n}=2\pi/\omega_{\ell n}\simeq 18$ (e.g., Turner 1973).

Figure~\ref{vecp} shows the vertical profiles of the first three anelastic
eigenvectors of degree $\ell = 1$ and radial orders $n=[0,\ 1,\ 2]$.
The associated (dimensionless) eigenvalues are
$\omega_{10} = 0.278,\ \omega_{11} = 0.184$ and $\omega_{12} = 0.133$. As
g-modes are evanescent in a convectively unstable layer, each eigenvector is trapped
in the bottom stably stratified zone and its amplitude rapidly decreases
in the convection zone, as observed for $z\le 1$
where both $\xi_x$ and $\xi_z$ stop oscillating and tend to zero.

Once the anelastic eigenvectors $\vec{\psi}_{\ell n}$ are computed,
we can determine the amplitudes of the g-modes propagating in our
simulation. Indeed, we have showed in Paper~I that these eigenvectors
are orthogonal to each other and form an orthogonal basis onto which the
simulated velocity field can be projected as

\beq
\hat{\vec{u}}_{\ell} (z,t) = \sum_{n=0}^{\infty} \langle\vec{\psi}_{\ell n},
\hat{\vec{u}}_{\ell}\rangle \vec{\psi}_{\ell n} + \hbox{``rest''},
\eeq
where the symbol $\langle\ ,\ \rangle$ denotes a scalar product defined by

\[
\langle \vec{\psi}_1,\vec{\psi}_2 \rangle = \int_{z_1}^{z_4}
\vec{\psi}^{\dagger}_1
\cdot \vec{\psi}_2\ \rho_0\ {\rm d}z,
\]
and the term ``rest'' contains all velocity components that are not due to
IGWs (e.g., acoustic waves, convective velocities, etc). Here
$\hat{\vec{u}}_{\ell}$ denotes the velocity field of degree $\ell$,
that is, the horizontal Fourier transform of the velocity field
$\vec{u}(x,z,t)$. We then define the amplitude of the g-mode of degree $\ell$
and order $n$ as the time-dependent complex coefficient

\beq
c_{\ell n} (t) = \langle \vec{\psi}_{\ell n},\hat{\vec{u}}_{\ell}
\rangle,
\eeqn{coeff}
which corresponds to the basis coefficient for the anelastic
eigenvector $\vec{\psi}_{\ell n}$.

\subsection{The evolution of the mode amplitudes}
\label{evol}

\begin{figure}
\centerline{\includegraphics[width=9cm]{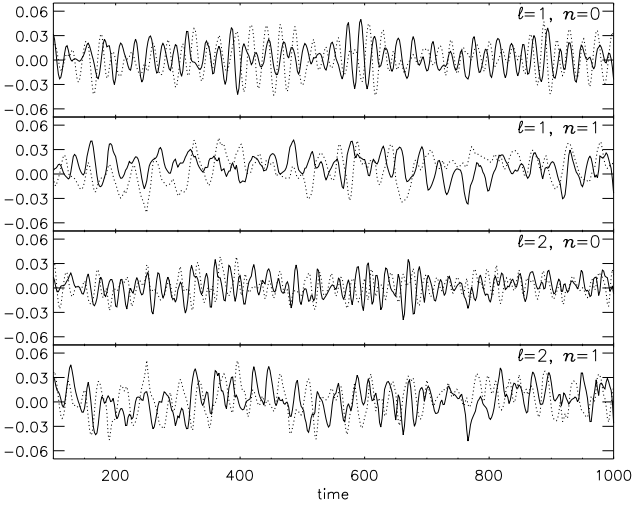}}

\caption{Evolution with time of the real (solid lines) and imaginary
(dotted lines) parts of the amplitude $c_{\ell n}$ of the g-modes of
degrees $\ell=[1,\ 2]$ and orders $n=[0,\ 1]$.}

\label{reA}
\end{figure}

\begin{figure}
\centerline{\includegraphics[width=8cm]{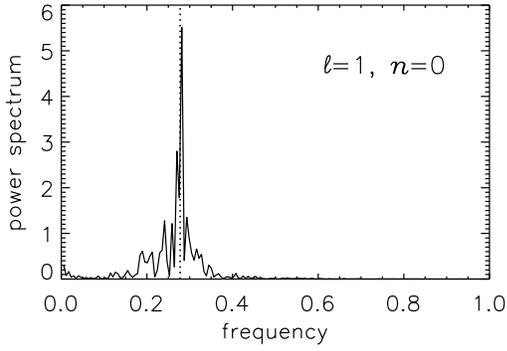}}

\caption{Corresponding temporal power spectrum of the mode amplitude
$c_{\ell n} (t)$ used in Fig.\ \ref{reA}, for $\ell=1$ and $n=0$. The
vertical dotted line marks the location of the theoretical frequency
$\omega_{10}$
of the underlying g-mode. Amplitudes have been magnified by $10^5$
for clarity.}

\label{spec1}
\end{figure}

We show in Fig.\ \ref{reA} the resulting amplitudes obtained by
applying our projection technique to
the numerical simulation. In this figure, we have plotted the real and
imaginary parts
of the complex amplitude \eq{coeff} for four g-modes of degrees $\ell
=[1,\ 2]$ and radial orders $n=[0,\ 1]$. When a standing gravity
wave occurs in a
hydrodynamical simulation, its complex amplitude $c_{\ell n}$ behaves as

\beq
c_{\ell n} \propto \exp(-\alpha t) \exp (\hbox{i}\omega_{\ell n} t),
\eeqn{law_ampli}
where $\omega_{\ell n}$ is the mode eigenfrequency and $\alpha$ a
coefficient proportional to the diffusion process. For instance in
the case of an excitation by an oscillating bubble dealt with in Paper~I,
the temporal Fourier transform of the mode amplitude leads to a single
peak centered around the eigenfrequency $\omega_{\ell n}$, with a width
that is proportional to the viscosity (see Fig.\ 6 in Paper~I). In other
words, each g-mode obeys in this situation the same law as that of
a linearly damped free harmonic oscillator.

However, in the case of a random excitation by penetrating plumes shown
in Fig.\ \ref{reA}, the real or imaginary parts now evolve either
chaotically around zero when the mode is not excited, or in a periodic
fashion as $\cos \omega_{\ell n} t$ or $\sin \omega_{\ell n} t$
when the mode is excited. Indeed, some wave events are well visible,
particularly for times $t=[400-600]$ in the $\ell=1,\ n=0$ diagram,
but the time evolution is mainly chaotic, suggesting that
such wave events are difficult to extract. As a consequence, taking
a temporal Fourier transform over the whole simulation of the mode
amplitude $c_{\ell n} (t)$ makes no sense as we will mix together
wave and non-wave events. This is illustrated in Fig.\ \ref{spec1} where
we computed the temporal Fourier transform of the mode amplitude in
Fig.\ \ref{reA}
for $\ell=1$ and $n=0$. Comparing to Paper~I, single peaks centered
around the theoretical eigenfrequencies $\omega_{\ell n}$ have been
replaced by a forest of peaks roughly centered around $\omega_{10}$,
i.e.\ the mixing of wave events with non-wave events degrades
the quality of the anelastic projection. Nevertheless, we remark that
the spectrum in Fig.\ \ref{spec1} is better than the one obtained in
the right-hand panel of Fig.\ \ref{classic} with the classical method as peaks are now
concentrated around the theoretical eigenfrequency. However, such spectra
do not permit a detailed study of the amplitudes of g-modes propagating
in the RZ, as other contributions (e.g., convective velocities in the
overshoot region as well as penetrations of downward plumes)
interfere with the time evolution of each mode amplitude.

\subsection{Time-frequency diagrams to extract wave events}

In order to accurately extract the hidden wave events, we use
time-frequency diagrams of the mode amplitude $c_{\ell n}(t)$, that is,
the temporal Fourier transforms
are computed by using a sliding window of fixed width
(e.g., Flandrin \& Stockler 1999). Assuming that this width is
$\Delta t$ (it is moreover beneficial to choose a multiple of the mode
period), we perform the following Fourier transform at time $t$

\[
\hat{c}_{\ell n} (t,\omega) = \int_{t-\Delta t/2}^{t+\Delta t/2}
c_{\ell n} (t')\ e^{{\rm i} \omega t'}\ {\rm dt'},
\]
and thus iterate the process at the next time $t+\delta t$, $\delta t$ being the
timestep of the simulation. We then obtain a 2-D representation of
the power spectrum $|\hat{c}_{\ell n} (\omega)|^2$ in a time-frequency
plane $(t,\omega)$ which highlights the time intervals during which the
corresponding g-mode is really excited in the RZ.

\begin{figure}
\centerline{\includegraphics[width=9cm]{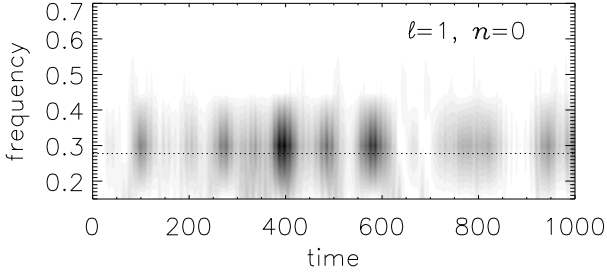}}

\caption{Time-frequency diagram of the amplitude of the g-mode at
$\ell=1$ and $n=0$ using a temporal window of width
$\Delta t = 2T_{10}$, with $T_{10}=2\pi/\omega_{10}
\simeq
22.6$ the mode period.
The horizontal dotted line corresponds to the mode
frequency $\omega_{10}$.}

\label{timefreq}
\end{figure}

\begin{figure}
\psfrag{toto}{Event function $\cal E$}
\centerline{\includegraphics[width=8.5cm]{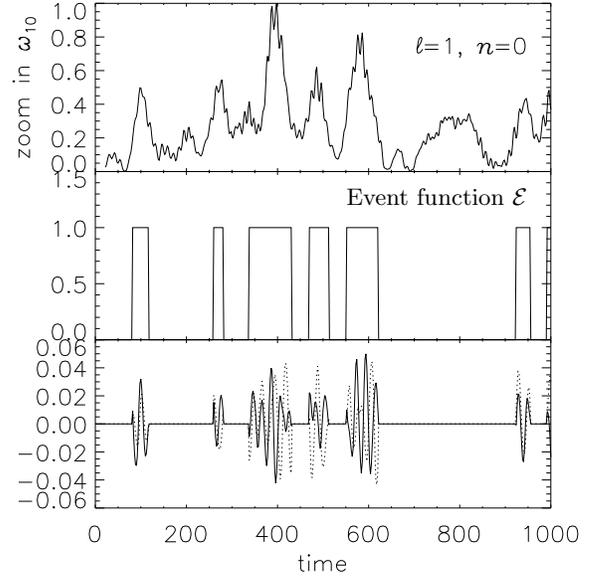}}

\caption{Filtering of the mode $\ell=1$, $n=0$:
zoom in Fig.\ \ref{timefreq} around the frequency $\omega_{10}$ (upper
panel); the resulting event function $\cal E$ given by \eq{fi} (middle
panel); the corresponding real (solid line) and imaginary (dotted line)
parts of the mode amplitude after the filtering (lower panel).}

\label{filter}
\end{figure}

To illustrate the utility of this method, we focus on the g-mode at
$\ell = 1$ and $n=0$ before applying it to other modes; see Fig.\
\ref{timefreq}.
Time intervals
during which this mode is excited to significant amplitudes are well extracted, as many bumps
appear along the line $\omega_{10}$, especially in the
range $t=[400-600]$ where three bumps are present. In order to isolate
precisely and {\it automatically} the most powerful wave events, we apply the
following procedure, illustrated in Fig.\ \ref{filter} still with our test
mode at $\ell=1$ and $n=0$:

\begin{itemize}

\item we first compute a mean profile of the time-frequency diagram around
the mode frequency $\omega_{\ell n}$ (shown in the upper panel).

\item we then build what we call the ``event function $\cal E$'', that
is, a function which is non-zero only when the previous mean profile
is higher than its mean value:

\greq
\hbox{if }\ f(t) <\ \langle f \rangle_t \ltex{then} {\cal E} =0, \\ \\
\hbox{if }\ f(t) \ge\ \langle f \rangle_t \ltex{then} {\cal E} =1,
\egreqn{fi}
and iterate the process four times by restarting from the new amplitude
profile ${\cal E}\times f$. We then obtain the final event
function $\cal E$ under a binary form (i.e.\ a succession of 0 and 1), shown in
the middle panel in Fig.\ \ref{filter}.
Six events are clearly isolated, four of them being clustered
in the range $t=[250-620]$.

\item finally, we apply the event function $\cal E$ to the
time-dependent amplitude $c_{\ell n}(t)$. It thus
emphasizes the time intervals during which $c_{\ell n} (t)$ behaves as
$\exp (\hbox{i} \omega_{\ell n} t)$, that is, during which the corresponding
g-mode is excited in the RZ; see the filtered real and imaginary
parts of $c_{10}$ in the lower panel of Fig.\ \ref{filter}.

\end{itemize}

\begin{figure}
\centerline{\includegraphics[width=8.5cm]{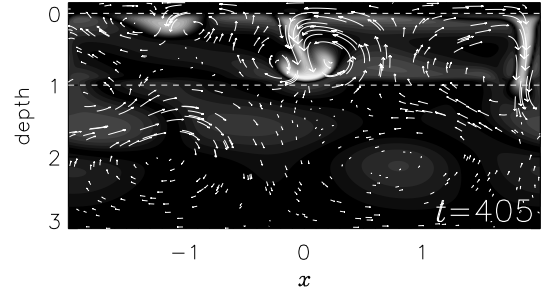}}

\caption{Snapshot of the velocity field superimposed on the internal
energy fluctuations for time $t=405$. Note the large-scale velocity field
in the bottom radiative zone mainly due to the standing g-mode at
$\ell=1$ and $n=0$.}

\label{showave}
\end{figure}

\begin{figure}
\centerline{\includegraphics[width=9.5cm]{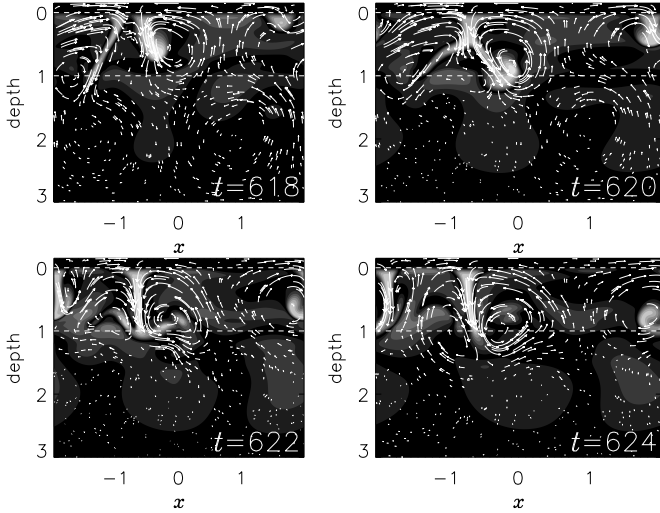}}

\caption{Destruction of the g-mode $\ell =1,\ n=0$ by a penetrating
plume entering in the RZ at time $t\simeq 620$.}

\label{dead}
\end{figure}

We obtain the longest wave event at $\ell=1$ and $n=0$ during times $t=[344-
435]$, which approximately corresponds to four mode periods, i.e.\
$\delta t = 91 \simeq
4 T_{10}$. This is confirmed in
the snapshot of the velocity field at time $t=405$ where a large-scale
velocity field, signature of the propagation of the g-mode at $\ell=1$
and $n=0$, is present in the bottom radiative zone (Fig.\
\ref{showave}). Now, why do modes die out? The
answer lies in Fig.\ \ref{dead}, where we plot four snapshots of the
velocity field for times $t=[618-624]$ which correspond to the end of
the events group clustered in the range $t=[250-620]$. The mode pattern
is simply
destroyed by the penetration of a plume which is born in the upper
part of the CZ at time $t\simeq 615$, then crosses the CZ and enters the RZ
at time
$t=620$ where it kills the mode propagation. As a consequence,
the large scale velocity field associated with the mode
disappears and the event
function $\cal E$ becomes zero.

\begin{figure}
\centerline{\includegraphics[width=8cm]{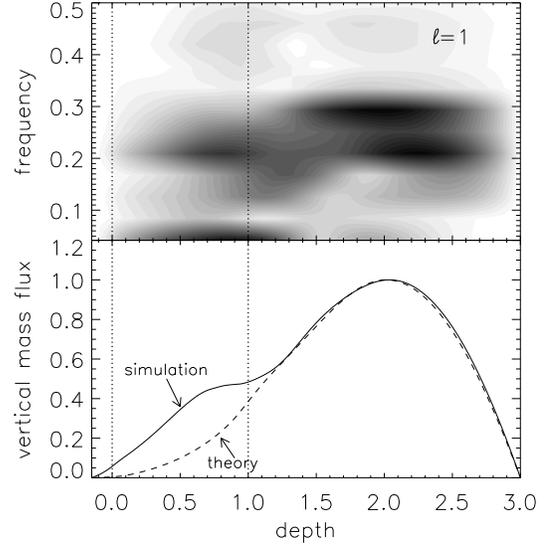}}

\caption{{\it Upper panel}: same as in Fig.~\ref{classic} for $\ell=1$
except that the temporal Fourier transform has been computed for time
$t=[300-450]$ only. {\it Lower panel}: comparison between the
corresponding vertical mass flux $\widehat{\rho w}$ (solid line) and the
theoretical
one (dashed line) computed from the anelastic eigenvector at $\ell =1$
and $n=0$.}

\label{shape}
\end{figure}

Before generalizing this formalism to more modes,
it is instructive to re-apply during this longest wave event the classical
method discussed in \S\ref{metclass}, in
order to compare both the spectrum and the simulated vertical mass flux
with the theoretical ones. This is what we did in Fig.\ \ref{shape},
which is the same as Fig.~\ref{classic} at $\ell = 1$ except that we
focus now on times $t=[300-450]$. As expected, the bump around
$\omega_{10}$ is more pronounced in the RZ (upper panel) and a comparison
between the shark fin vertical profile deduced from a zoom around
$\omega_{10}$ with the one computed from the theoretical anelastic mode
shows an almost perfect agreement there (lower panel). It means that
the dynamics of this region is well reproduced by our linear anelastic
model. On the contrary, as g-modes cannot propagate in an unstable layer,
it is normal to find a large discrepancy between these two profiles in
the convection zone.

\begin{figure}
\centerline{\includegraphics[width=9cm]{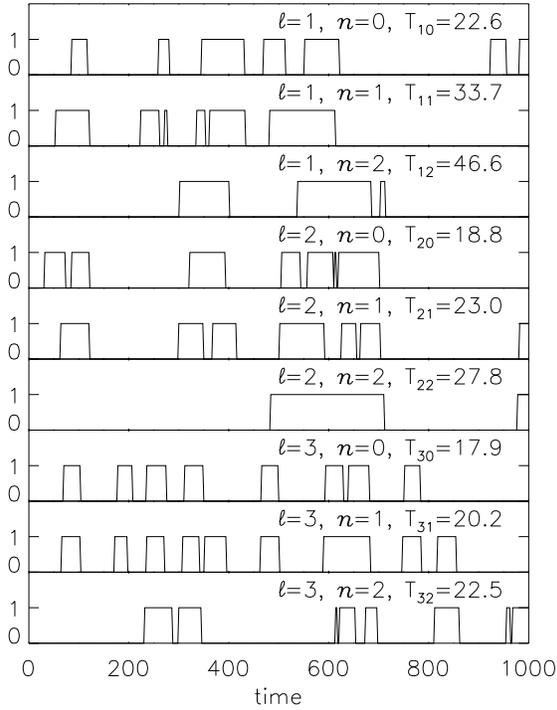}}

\caption{The event functions $\cal E$ for g-modes of degrees
$\ell=[1,2,3]$ and radial orders $n=[0,1,2]$, with their associated
periods $T_{\ell n}$.}

\label{events}
\end{figure}

\begin{figure}
\centerline{\includegraphics[width=9cm]{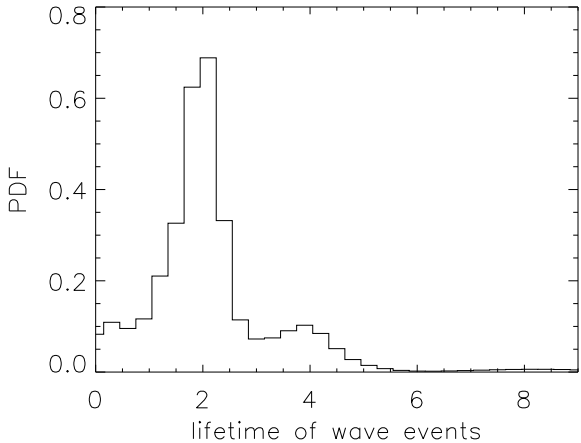}}

\caption{PDF of the lifetime of wave events, in units of the
mode periods.}

\label{pdf}
\end{figure}

We then apply this method to the g-modes of the first three degrees $\ell =
[1,\ 2,\ 3]$ and radial orders $n=[0,\ 1,\ 2]$, with resulting event functions
$\cal E$ given in Fig.\ \ref{events}. That allows us first to show that
the second bump located
around $\omega \sim 0.2$ in the upper panel in Fig.\ \ref{shape} is due
to the propagation of the mode $\ell=1,\ n=1$, as the corresponding event
function is not zero for times $t \simeq 400$.
Second, the assembly of the whole event function permits us to perform a
statistical study of the mode lifetimes, that is, to compute the PDF of the
duration of events (Fig.~\ref{pdf}). This PDF is peaked around
2, meaning that the lifetime of a mode is approximately twice its
period. Compared to the solar acoustic modes for which
$T_{\hbox{\scriptsize osc}} \simeq$ 5min and $T_{\hbox{\scriptsize
life}} \sim\hbox{ hour}\sim 20T_{\hbox{\scriptsize osc}}$, such a ratio is
very small, i.e. g-modes patterns are rapidly destroyed by the following
penetrating plumes and it may be a problem for their detection at the
star surface.

\subsection{Kinetic energy due to g-modes}

Using the previous time-frequency diagrams for every g-mode allows
us to find the time intervals during which IGWs propagate
in the RZ. Now we want to quantify the efficiency of this
stochastic excitation by using, say, some wave flux in the vertical
direction. However, as we impose the reflective boundary condition $w=0$
for the vertical velocity
both at the surface $z=z_1$ and the bottom $z=z_4$ of our computational
domain, we have standing waves rather than propagating waves
and no flux is carried by waves in the vertical direction. As in Paper
I, we thus chose to focus on the kinetic energy embedded in g-modes
using the following Parseval type relation valid in the anelastic
subspace

\beq
\int_V \rho \vec{u}^2 (x,z)\ {\rm d}x\ {\rm d}z =
\sum^{+\infty}_{\ell,\ n} |c_{\ell n}|^2 + ``{\rm rest}",
\eeqn{parseval}
where the left-hand side corresponds to the total kinetic energy in
the simulation at a given time $t$ and the term ``rest" contains all
the contributions which are not due to IGWs (the demonstration is given
in Paper I). This relation is very useful as it allows to quantify the
kinetic energy embedded in every g-mode $(\ell,\ n)$. Indeed, using
the classical Parseval relation just allows one to quantify
the amount of (say) kinetic energy contained in a mode of given degree
$\ell$ and contributions coming from g-modes as well as p-modes or any
turbulent motion are mixed together. The advantage of our anelastic
subspace relation \eq{parseval} is that it lifts this degeneracy in
the radial order $n$ by isolating the g-mode contributions. Of course,
this relation should be applied only during times when IGWs effectively
propagate, i.e.\ when $c_{\ell n} \propto \exp (\rm{i} \omega_{\ell n}
t)$ or ${\cal E} = 1$, such that the $|c_{\ell n}|^2$ contributions
make sense.

\begin{figure}
\centerline{\includegraphics[width=8.5cm]{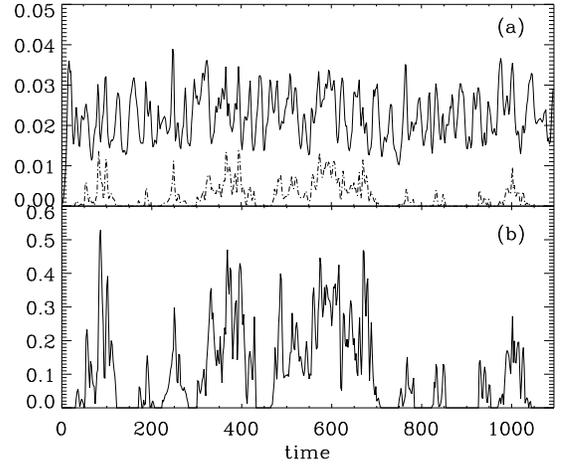}}

\caption{{\bf a):} time evolution of the total kinetic energy
(solid line) embedded in the simulation, with its component which is
only due to g-modes (dot-dashed
line). {\bf b):} ratio between the two.}

\label{etot}
\end{figure}

The temporal evolution of the total kinetic energy $E_{\hbox{\scriptsize
tot}}$ embedded in the
simulation (i.e.\ the left-hand
side in Eq.~\ref{parseval}) is illustrated
as the solid line in Fig.\ \ref{etot}a, while the
dot-dashed line in the same panel corresponds to the contribution
$E_{\hbox{\scriptsize IGW}}$ coming
from g-modes only (the right-hand side sum in Eq.~\ref{parseval},
where the $\ell=[1-3]$ and $n=[0-2]$ modes have been considered).
The interesting quantity is of course the ratio between the two, that is
$E_{\hbox{\scriptsize IGW}}/E_{\hbox{\scriptsize tot}}$, which emphasizes
the efficiency of the IGW excitation by the downward plumes (Fig.\
\ref{etot}b). It emerges that g-modes contribute up to about 40\% of the
total kinetic energy when they are excited, for example in the range $t\simeq
[400-600]$ when the $l=1,\ n=0$ mode is strongly excited.
This large ratio is interesting, since it means that internal gravity waves may
contain a non-negligible part of the total kinetic energy of the coupled
system formed by the two neighboring convective and radiative zones.
This is a direct demonstration that the excitation of IGWs by penetration
plumes can be quite efficient, at least in 2-D.

\section{Conclusions}
\label{conclu}

We have investigated numerically the excitation of internal
gravity waves by the penetration of convective plumes into an adjacent
stably stratified zone. This problem is intimately related to the
transport processes of chemicals and/or angular momentum in radiative
zones of cool stars, such as the Sun. The knowledge of both
spectrum and amplitudes of the internal waves field is crucial.

After recalling the main features of 2-D hydrodynamical simulations
of penetrative convection, we focused on the problem of the mode
identification by first using a classical method based on successive
Fourier transforms of the vertical mass flux over the whole simulation. We
thus showed that this tool is not adapted to detect the g-modes
propagating in these simulations as the resulting spectra are very noisy,
preventing a quantitative study of the phenomenon.

We then introduced our method for detecting accurately the internal waves
in the radiative zone. It is based on the projection of the simulated
velocity field onto the anelastic eigenmodes that are solutions of the
associated linear eigenvalue problem for the perturbations. Indeed, when
a standing wave is present near the bottom radiative zone, its spatial
structure is that of an eigenmode and the coefficient of the projection
onto this eigenmode gives the mode amplitude. This amplitude is of
course time-dependent as the internal wave is only generated after the
penetration of plumes, and is then dissipated under the action of both
viscous and radiative diffusions. This leads us to use time-frequency
diagrams to isolate the most powerful wave events and to construct what
we called ``the event function $\cal E$'', that is, a binary function
(0/1) which is set to 1 only when the mode amplitude is higher than a
given threshold. As an example, we focused on the g-mode at $\ell =1$
and $n=0$, whose period is $T_{10} = 22.6$. We extracted six wave
events in our particular 2-D simulation, the longest one corresponding to
four mode periods. We then showed the intricate link existing between the
mode and the downward plumes as they can either excite it or destroy it!

The extension of this study to the g-modes at $\ell=[1-3]$ and $n=[0-2]$
allowed us to compute the PDF of the mode lifetimes. We found that
the mean lifetime is only around twice the period of the mode.
The shortness of this
lifetime may be a problem from an observational point of view where one
needs lifetimes as big as possible (the large-scale solar acoustic modes
have lifetimes of about the day, i.e.\ several hundreds of
times
their mean period).

Finally, we looked at the kinetic energy content of the excited g-modes
and showed that up to 40\% of the total kinetic energy at times may
reside in g-modes. This level is only reached during a fraction of
the time, and the mode kinetic energy varies considerable with time.
Nevertheless, when the modes are excited, the corresponding
velocity field in the radiative zone has an amplitude that may
lead to an efficient wave transport there (through the advective term
$\vec{u}_{\hbox{\scriptsize wave}}\cdot \na$).

It is clear that our detection method allows a {\it quantitative} analysis
of the problem of g-mode excitation by penetrative convection.
Following this work, we have been doing a parametric study of the influence of
the convective flux on the mode amplitudes, by trying to predict these
amplitudes from
mixing-length arguments. Some recent 2-D simulations by Kiraga et al.\
(2003) indeed suggest that such mixing length models systematically underestimate
the strength of the internal wave field.

In the same way, the depth of the penetration in the stably stratified
zone is without doubt a key parameter in the excitation mechanism by
penetrative plumes. This penetration strongly depends on the local value
of the P\'eclet number Pe: {\it (i)} large values of Pe mean that the advection
is greater than the radiative diffusion such that the plume keeps its
identity and is stopped very rapidly by the buoyancy braking, leading to
a tiny penetration; {\it (ii)} small values of Pe mean that the plume thermalizes
very rapidly with its surrounding and the buoyancy braking disappears,
leading to a large penetration. It would then be interesting to further
study the influence of the P\'eclet number on the mode amplitudes, by
computing for example a grid of 2-D polytropic models with different
indexes $m_3$, in order to play with the P\'eclet number jump at the
base of the convection zone (Eq.\ \ref{pe_jump}).
Likewise, the influence of the dimensionality; i.e.\ the differences
between 2-D and 3-D also need be investigated.

\acknowledgements
This work has been supported by the European Commission under Marie-Curie
grant No.\ HPMF-CT-1999-00411. Calculations were carried out on the
CalMip machine of the ``Centre Interuniversitaire de Toulouse'' (CICT)
which is acknowledged. We thank the referee (J.-P. Zahn) for his
meaningful comments.

\appendix

\section{The initial stratification in the three layers}
\label{init}

The initial vertical stratification is computed using polytropes of
various indexes for which

\[
P \propto \rho^{1+1/m} \ltex{or} \rho \propto T^m,
\]
where $m$ denotes the polytropic index.  Such polytropic solutions
are useful in numerical simulations of convection as they allow to
easily specify the stability (or not) of a layer to convection. Indeed,
the well known Schwarzschild criteria for convection are (e.g., Hansen \&
Kawaler 1994)

\greq
\del < \delad \Rightarrow \hbox{STABLE}, \\ \\
\del > \delad \Rightarrow \hbox{UNSTABLE},
\egreq
where $\del\equiv d\ln T/d\ln P$ and $\delad=1-1/\gamma$ is its value
in the case of an adiabatic stratification. The using of polytropic
solutions leads to the following simple form for the ``del"

\[
\del = \frac{1}{1+m},
\]
meaning that a polytropic layer of index $m$ will be stable or unstable
to convection following

\greq
\del < \delad \Rightarrow m > \mad\ \hbox{: STABLE}, \\ \\
\del > \delad \Rightarrow m < \mad\ \hbox{: UNSTABLE}.
\egreq
Since $\gamma = 5/3$, we have $\delad =
2/5$ and $\mad = 3/2$ in this case and a polytropic layer will be
convectively stable if $m>3/2$ and unstable if $m<3/2$.

\begin{figure}
\centerline{\includegraphics[width=7cm]{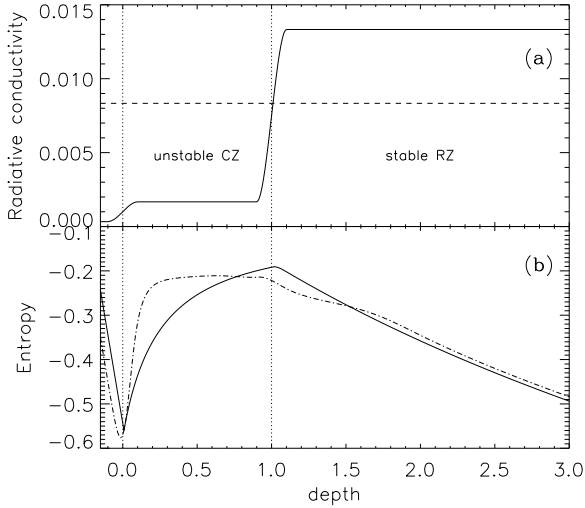}}

 \caption{Initial vertical profiles for a layered system with
 $m=[-0.9,-0.5,3]$ of the radiative conductivity (a) and entropy
 (b). The vertical dotted lines delimit the convection zone, whereas the
 horizontal dashed line in the (a)-plot corresponds to the critical value
 ${\cal K}_{\hbox{\scriptsize ad}}$ below which the layer is convectively
 unstable. The dot-dashed line in the (b)-plot corresponds to the entropy
 profile in the relaxed state.}

\label{strat}
\end{figure}

Once the three polytropic indexes of the superposed layers have been fixed
we first compute the corresponding radiative conductivity profile by
assuming that all of the energy flux $\fb$ that we impose at the bottom is
initially transported by radiation, that is,

\beq
{\cal K}_i = \frac{\fb}{g} (\gamma -1 ) (1+m_i),
\eeqn{rad0}
in the layer $i$ ($i=[1,2,3]$). In this formalism, the radiative
conductivity is thus a constant in each layer and its three different
values are joined using polynomials of the third-order.

In Fig.~\ref{strat}a we show an example of such a profile for a polytropic
layered system with indexes $m_1=-0.9,\ m_2=-0.5$ and $m_3=3$. As expected,
the radiative conductivity in the CZ is below the critical value ${\cal
K}_{\hbox{\scriptsize ad}}$ given by

\[
{\cal K}_{\hbox{\scriptsize ad}} = \frac{\fb}{g} (\gamma -1 ) (1+\mad),
\]
that is, the value deduced from Eq.~\eq{rad0} using $m=\mad$. More
surprising could be the value ${\cal K}_1 < {\cal K}_{\hbox{\scriptsize
ad}}$ in the surface layer as one rather expects a stably stratified layer
here, as observed in solar-like stars where a very thin superficial stable
layer exists. In fact, this layer is stable due to its efficient cooling
[Eq.~(\ref{cooling})] such that the Schwarzschild criterion does not
apply. We thus adopt a very weak value for ${\cal K}_1$, by taking
$m_1=-0.9$ in all of the simulations, and compute the initial hydrostatic
stratification of this layer by assuming that it is already isothermal
with ${\rm d}e/{\rm d}z = 0$.

Concerning the initial stratification of the CZ, things are more
involved. Assuming an initial polytropic stratification with $\del_2 =
1/(1+m_2)$ is clearly not a good idea, as an efficient convection is
always associated with an almost adiabatic stratification with
$\del_2 \simeq \delad$. In this case, the relaxation of the CZ towards
its adiabatic state would take a lot of numerical timesteps. One
simple solution of this problem consists in starting from an adiabatic
stratification in the CZ by imposing $\del_2 = \delad$. However, this
solution does not take into account the entropy jump which appears at
the top of the CZ (the difference between constant entropy in the CZ and
the entropy minimum in the photosphere; see e.g., Abbett et al.\ 1997;
Ludwig, Freytag \& Steffen 1999) such that the relaxation time would
stay important.

One solution of this relaxation time problem consists in starting from
a mixing length stratification where the (local) superadiabatic gradient
in the CZ is modeled using the following mixing-length argument

\[
\del^{\hbox{\scriptsize mlt}}_2 = \delad + \alpha \lp
\frac{\fc}{\rho c^3_s} \rp^{2/3},
\]
where $c_s=\sqrt{\gamma(\gamma-1)e}$ denotes the local sound speed, $\alpha
\simeq 1.5$ is a free mixing-length parameter and $\fc$ is the convective
flux given by

\[
\fc = \fb - \fr = \fb \left[ 1 - \frac{1}{g}(\gamma -1)(1+m_2) \right].
\]
Finally, the initial stratification of the bottom RZ is simply computed
by assuming that all of the bottom flux is transported in this layer by
radiation and one sets $\del^{\hbox{\scriptsize rad}}_3 = 1/(1+m_3)$.

To summarize, the initial stratification of our three-layer model is
computed from

\[
\dnz{\ln \rho} = \frac{1}{\gamma -1} \frac{g}{e},
\]
where $e$ obeys to the following set of equations in the three layers

\[ \left\{ \begin{array}{ll}
\disp \dnz{e} = 0 & \ltex{if} z_1\le z \le z_2, \\ \\
\disp \dnz{e} = \frac{g}{\gamma -1} \del^{\hbox{\scriptsize mlt}}_2
& \ltex{if} z_2< z < z_3, \\ \\
\disp \dnz{e} = \frac{g}{\gamma -1} \del^{\hbox{\scriptsize rad}}_3
& \ltex{if} z_3\le z \le z_4.
\end{array} \right.
\]
and these differential equations are iterated until the density at the
BCZ is equal to 1; i.e.\ we impose $\rho (z=z_3) =1$.

In Fig.~\ref{strat}b we show the resulting vertical profile of the initial
entropy $s/c_p = \ln (p/\rho^{\gamma})/\gamma$ for the same
indexes $m=[-0.9,-0.5,3.]$. The solid line denotes the
initial entropy whereas the dot-dashed line corresponds to its profile
in the relaxed state. First, as the entropy gradient in an isothermal
layer is a constant, one verifies that $s \propto z$ for $z=[z_1,z_2]$.
Second, the comparison between the initial and relaxed profiles shows
that the mixing-length stratification well reproduces the entropy jump at
the top of the CZ: the strong mixing taking place in the deep layers of
the CZ leads to an almost flat entropy profile, which disappears at the
base of the photosphere. As a consequence, the computing time needed to
relax towards this solution is considerably reduced by using mixing-length
solutions.


\begin{thebibliography}{}
\bibitem{} Abbett, W.~P., Beaver, M., Davids, B., Georgobiani, D.,
Rathbun, P., \& Stein, R.~F. 1997, ApJ, 480, 395
\bibitem{} Alexander, M.~J. \& Pfister, L. 1995, Geophys.~Res.~Lett., 22, 2029
\bibitem{} Bogdan, T.~J., Cattaneo, F., \& Malagoli, A. 1993, ApJ, 407, 316
\bibitem{} Brandenburg, A., Jennings, R.~L., Nordlund, \AA., Rieutord,
M., Stein, R.~F., \& Tuominen, I. 1996, J.~Fluid~Mech., 306, 325
\bibitem{} Bretherton, F.~P. 1969, Quart.~J.~Roy.~Meteo.~Soc., 95, 213
\bibitem{} Brown, T.~M., Christensen-Dalsgaard, J., Dziembowski, W.~A.,
Goode, P., \& Gough, D.~O. 1989, ApJ, 343, 526
\bibitem{} Brummell, N.~H., Clune, T.~L., \& Toomre, J. 2002, ApJ, 570,
825
\bibitem{} Cattaneo, F., Brummell, N.~H., Toomre, J., Malagoli, A., \&
Hurlburt, N.~E. 1991, ApJ, 370, 282
\bibitem{} Dintrans, B. \& Brandenburg, A. 2004, A\&A, 421, 775 (Paper~I)
\bibitem{} Dintrans, B., \& Rieutord, M. 2001, MNRAS, 324, 635
\bibitem{} Flandrin, P., \& Stockler, J. 1999, Time-Frequency/Time-Scale
Analysis ({Academic Press})
\bibitem{} Gabriel, A.~H., Baudin, F., Boumier, P., et al. 2002,
A\&A, 390, 1119
\bibitem{} Gough, D.~O., Moore, D.~R., Spiegel, E.~A., et al. 1976,
ApJ, 206, 536
\bibitem{} Graham, E. 1975, J.~Fluid~Mech., 70, 689
\bibitem{} Hansen, C.~J., \& Kawaler, S.D. 1994, Stellar interiors
(New-York: Springer-Verlag)
\bibitem{} Hurlburt, N.~E., Toomre, J., \& Massaguer, J.~M. 1984, ApJ,
282, 557
\bibitem{} Hurlburt, N.~E., Toomre, J., \& Massaguer, J.~M. 1986, ApJ,
311, 563
\bibitem{} Hurlburt, N.~E., Toomre, J., Massaguer, J.~M., \& Zahn,
J.-P. 1994, ApJ, 421, 245
\bibitem{} Hyman, J.\ M. 1979, in Adv. in Computation Methods for
Partial Differential Equations, Vol. III, ed. R. Vichnevetsky, \&
R.~S. Stepleman, Publ. IMACS, 313
\bibitem{} Iben, I.~J. 1965, ApJ, 142, 1447
\bibitem{} Kiraga, M., Jahn, K., St{\c e}pie{\' n}, K. \& Zahn, J.-P.
2003, Acta Astron., 53, 321
\bibitem{} Kumar, P. \& Quataert, E.~J. 1997, ApJ, 475, L143
\bibitem{} Kumar, P., Talon, S., \& Zahn, J.-P. 1999, ApJ, 520, 859
\bibitem{} Landau, L., \& Lifshitz, E. 1980, Fluid Mechanics
(Oxford: Pergamon Press)
\bibitem{} Lele, S.~K. 1992, J.\ Comp.\ Phys., 103, 16
\bibitem{} Ludwig, H.-G., Freytag, B. \& Steffen, M. 1999, A\&A, 346,
111
\bibitem{} MacGregor, K.~B. \& Charbonneau, P. 1999, ApJ, 519, 911
\bibitem{} Nordlund, \AA., \& Stein, R.~F. 1990, Comput.~Phys.~Commun.,
59, 119
\bibitem{} Rieutord, M. \& Dintrans, B. 2002, MNRAS, 337, 1087
\bibitem{} Schatzman, E. 1993, A\&A, 279, 431
\bibitem{} Talon, S. \& Charbonnel, C. 1998, A\&A, 335, 959
\bibitem{} Talon, S., Kumar, P., \& Zahn, J.-P. 2002, ApJ, 574,
L175
\bibitem{} Turck-Chi\`eze, S., Garc{\'{\i}}a, R.~A., Couvidat, S.,
et al. 2004, ApJ, 604, 455
\bibitem{} Turner, J.~S. 1973, Buoyancy effects in fluids
(New-York: Cambridge University Press)
\bibitem{} Zahn, J.-P. 1991, A\&A, 252, 179
\bibitem{} Zahn, J.-P. 1994, A\&A, 288, 829
\bibitem{} Zahn, J.-P., Talon, S., \& Matias, J. 1997, A\&A, 322, 320
\end{thebibliography}
\end{document}